\documentclass[aip, apl, amsmath, amssymb, reprint]{revtex4-1}
\pdfoutput=1
\usepackage{graphicx}
\usepackage{dcolumn}
\usepackage{bm}

\begin{document}%

\title{Detection of Nanoparticles with a Frequency Locked Whispering Gallery Mode Microresonator}%

\author{Jon D. Swaim}%
 \email{jswaim@physics.uq.edu.au}%
\affiliation{Department of Physics, University of Queensland, St Lucia, QLD 4072 Australia}%
\author{Joachim Knittel}%
\affiliation{Department of Physics, University of Queensland, St Lucia, QLD 4072 Australia}%
\author{Warwick P. Bowen}%
\affiliation{Department of Physics, University of Queensland, St Lucia, QLD 4072 Australia}%
\affiliation{Centre for Engineered Quantum Systems, University of Queensland, St Lucia, QLD 4072, Australia}%

\date{\today}%

\begin{abstract}%
We detect 39 nm$\times$10 nm gold nanorods using a microtoroid stabilized via the Pound-Drever-Hall method.  Real-time detection is achieved with signal-to-noise ratios up to 12.2.  These nanoparticles are a factor of three smaller in volume than any other nanoparticle detected using WGM sensing to date.  We show through repeated experiments that the measurements are reliable, and verify the presence of single nanorods on the microtoroid surface using electron microscopy.  At our current noise level, the plasmonic enhancement of these nanorods could enable detection of proteins with radii as small as $a$ = 2 nm.
\end{abstract}%

\maketitle%

Label-free single molecule detection has been an active area of research in optics during recent years, in part due to the emergence of whispering gallery mode (WGM) resonators as ultra-sensitive refractive index sensors~\cite{Armani, Arnold}.  Recently, however, several theoretical studies have emphasized that the predicted detection limit of current devices is well above that which is required for single molecule sensitivity~\cite{Swaim}.  For this reason many efforts have been made to improve the signal-to-noise ratio (SNR) and thereby reach the single molecule limit, including interferometry~\cite{Lu, Knittel}, plasmonic enhancement~\cite{Shopova, Vollmer, Swaim, Dantham} and frequency stabilization~\cite{DOShea, Stern, He}.  In this letter we build on these works, demonstrating real-time detection of gold (Au) nanorods with a silica microtoroid stabilized using the Pound-Drever-Hall (PDH) technique~\cite{PDH}.  We detect 39 nm$\times$10 nm nanorods with a SNR up to 12.2 and a resonator quality (Q) factor of 6$\times$10$^5$.  These nanoparticles are $\sim$3 times smaller in volume than the smallest nanoparticles detected to date using the WGM sensing principle~\cite{Lu}.

The essence of the PDH stabilization technique is in the measurement of an error signal which is fed back into the laser to supress fluctuations in frequency.  The advantage of the technique is that it utilizes nulled lock-in detection, and the error signal is insensitive to a amplitude noise from the laser~\cite{PDH}.  Because the laser is stabilized with respect to a reference cavity (in our case, a microtoroidal WGM resonator), the feedback loop ensures that the laser's frequency will follow any frequency shift $\delta \omega$ in the cavity resonance, such as that experienced when a molecule binds to the microtoroid surface~\cite{Arnold, Swaim}.  

\begin{figure}[ht!]
	\begin{center}
	\includegraphics*[scale=0.5]{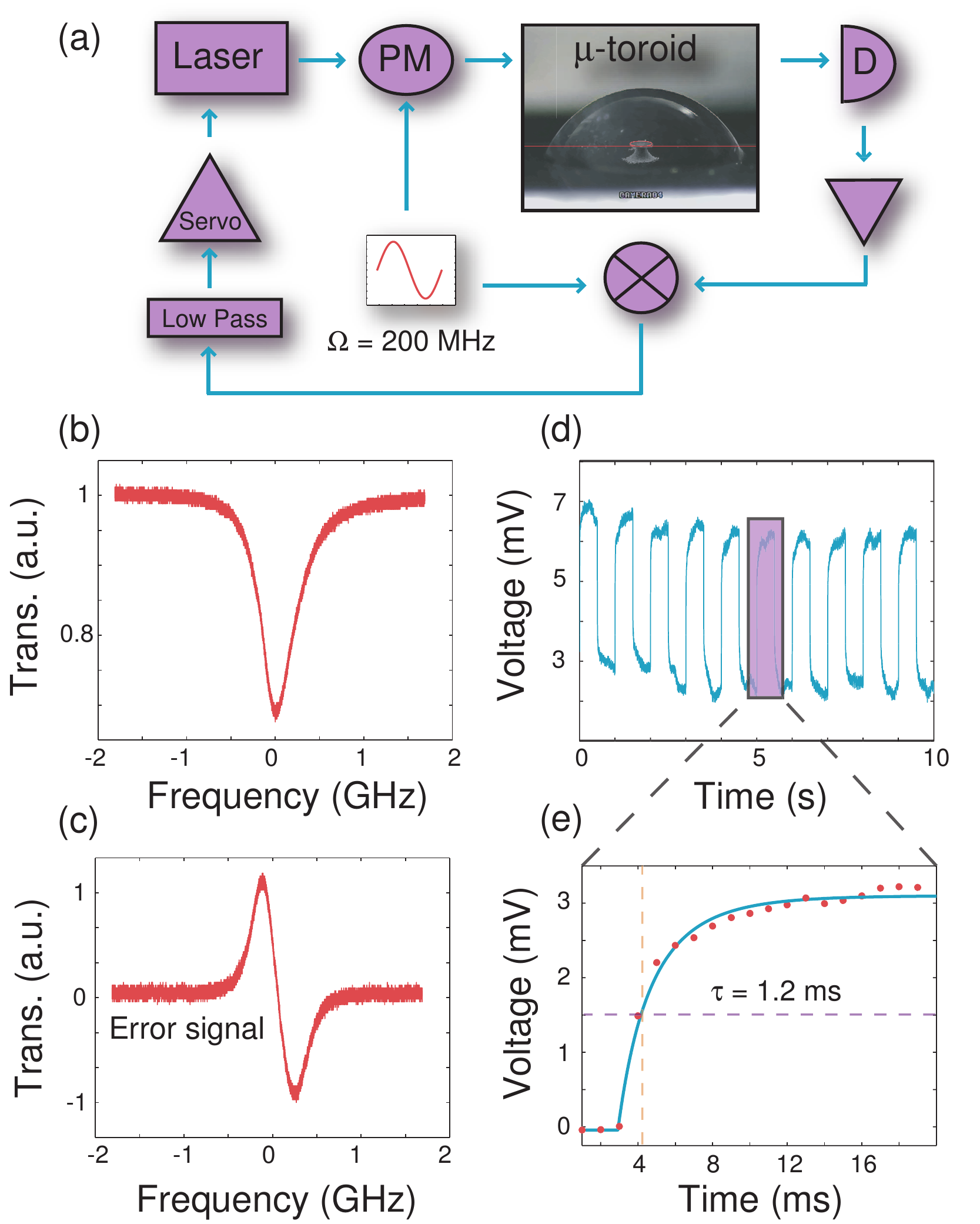}
	\caption{(a) Experimental schematic of the microtoroid immersed in water and stabilized via the PDH method.  (b) Microtoroid resonance (linewidth of $\gamma$ = 451 MHz) in water.  (c) PDH error signal for $\Omega$ = 200 MHz. (d) Feedback signal in response to a 1 Hz frequency modulation.  (e) Transient response of the feedback loop.  The fit reveals a 3 dB time constant of $\tau$ = 1.2 ms}
	\label{fig:setup}
	\end{center}
\end{figure}

Fig.~\ref{fig:setup}a shows a schematic illustrating the experimental setup.  The setup consists of a 780 nm laser source (New Focus 6300-LN) coupled to a LiNbO$_3$ phase modulator (PM) and then to a silica microtoroid via a tapered optical fiber.  A typical transmission spectrum of a microtoroid resonance in water is shown in Fig.~\ref{fig:setup}b.  The transmitted light is sent to a photodetector (D) and the output photocurrent is mixed with a local oscillator ($\Omega$ = 200 MHz) which drives the phase modulator.  The mixer output is sent through a low-pass filter to isolate the error signal (Fig.~\ref{fig:setup}c) and then to a high-speed proportional-integrator controller to supply the laser with a frequency feedback voltage. 

\begin{figure*}[ht!]
	\begin{center}
	\includegraphics*[scale=0.95]{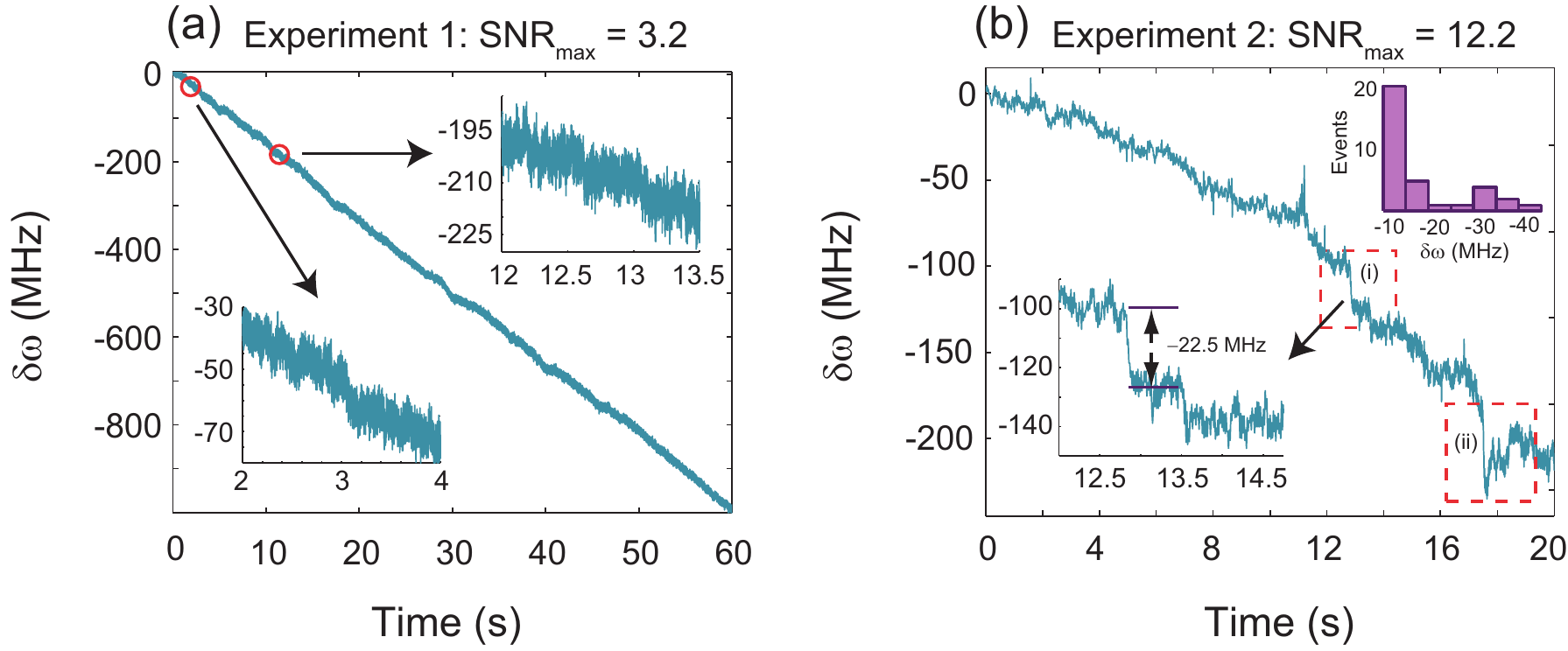}
	\caption{(a) PDH feedback signal for the first experiment.  The left and right insets magnify binding events around 2 s and 12 s, respectively.  (b)  PDH feedback signal for the second experiment.  The left inset shows the binding event at 13 s magnified, and the right inset shows a histogram of frequency shifts determined from the step-finder algorithm (bin size of 5 MHz).}
	\label{fig:data}
	\end{center}
\end{figure*} 

To emulate single nanoparticle binding events, we applied a 1 Hz frequency modulation on the light and then measured the locking feedback voltage.  This can be converted to an effective frequency shift $\delta \omega$ via calibration.  As expected, this resulted in a step-wise adjustment to the feedback signal (Fig.~\ref{fig:setup}d).  In Fig.~\ref{fig:setup}e we show the transient response of the feedback loop, with a measured time constant of $\tau$ = 1.2 ms.

We applied the PDH method to detect Au nanorods in two separate experiments, achieving qualitatively similar results.  In these experiments, the microtoroid was coupled to the tapered fiber and then immersed in a droplet of water as shown in Fig.~\ref{fig:setup}a.  The laser frequency was then locked to the microtoroid resonance.  The position of the tapered fiber was precisely controlled using a piezo-actuated stage, so as to avoid contacting the taper with the microtoroid.   Once the feedback signal was stable, a dilute solution of Au nanorods stabilized with cetyl trimethylammonium bromide (CTAB) was added to the droplet, and the PDH feedback signal was monitored on an oscilloscope.  Due to absorption by water at 780 nm, the resonator Q factors in the first and second experiments were 6.6$\times$10$^5$ and 6$\times$10$^5$, respectively.  The shape of the error signal shown in Fig.~\ref{fig:setup}c results from the fact that the cavity linewidth $\gamma > \Omega$. 

\begin{figure*}[ht!]
	\begin{center}
	\includegraphics*[scale=0.75]{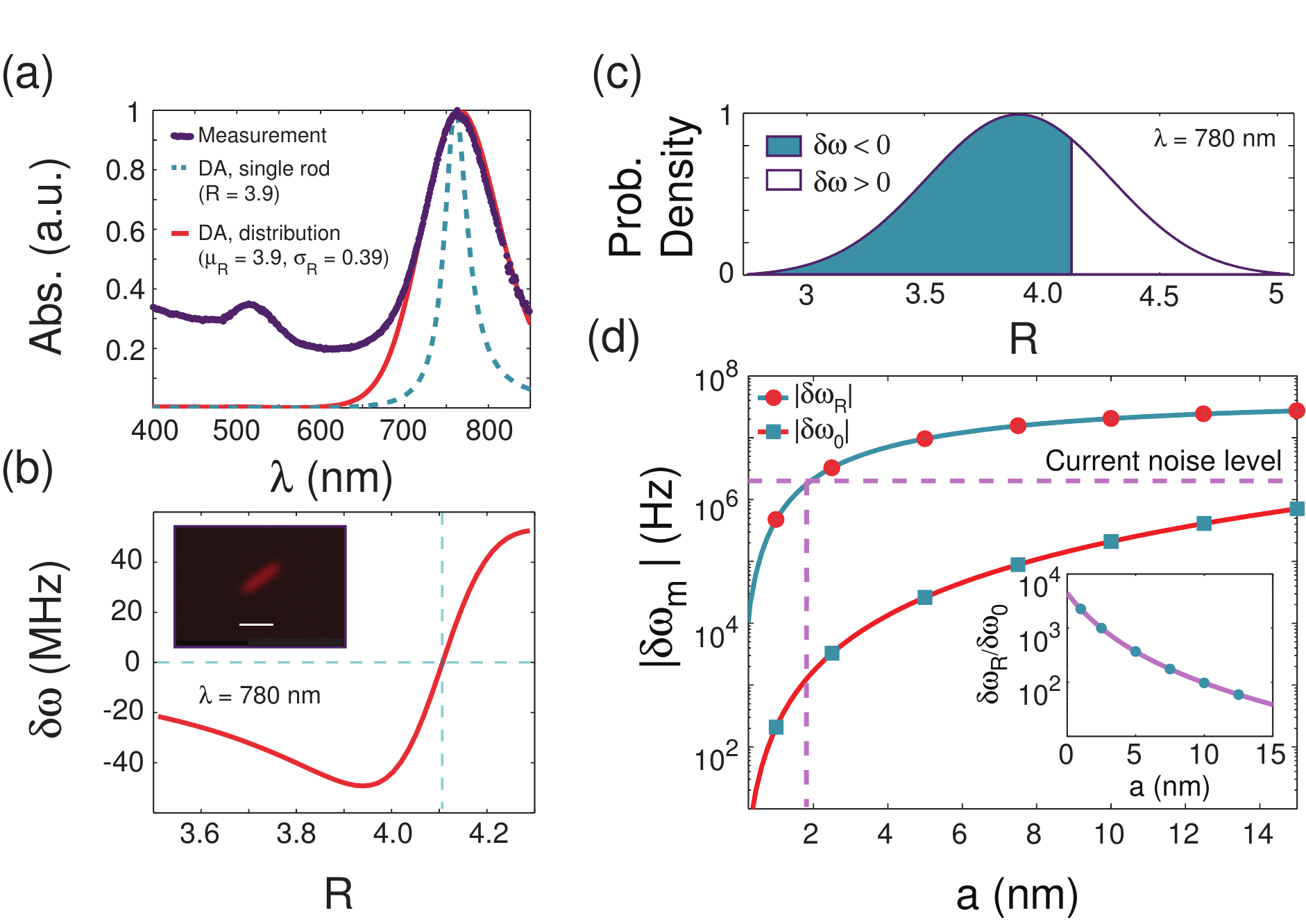}
	\caption{(a) Comparison of measured absorption spectra with theoretical values predicted in the dipole approximation (DA).  (b)  Predicted frequency shift as a function of aspect ratio $R$.  A false-colored SEM image of a nanorod bound to a microtoroid surface is shown in the inset (scale bar is 40 nm). (c) Probability density distribution of nanorod aspect ratios and corresponding sign of induced frequency shift. (d) Predicted single molecule frequency shift with and without plasmonic enhancement versus molecule radius $a$.  The inset shows the nanorod enhancement factor.}
	\label{fig:nanorods}
	\end{center}
\end{figure*}  

In Fig.~\ref{fig:data}, we show the measured feedback signals for the two experiments upon addition of the Au nanorods.  Frequency drifts of -16.5 MHz/s and -6.5 MHz/s are clearly seen in these data, as shown in other experiments~\cite{Lu, Arnold}.  In addition, we observed two types of nanorod interactions: (i) frequency shifts corresponding to binding events and (ii) frequency shifts followed closely by an oscillation (over a duration of 1-2 s) and subsequent escape of the nanorod.  Many of these type (i) shifts were observed in the first experiment (e.g., at 3 s, 12.2 s, 12.6 s and 13.1 s in the insets of Fig.~\ref{fig:data}a), as well as the second experiment (left inset of Fig.~\ref{fig:data}b at 13 s and 13.5 s).  Also, several of the type (ii) interaction were observed (e.g., at 17 s in Fig.~\ref{fig:data}b).  We believe these events arise from the electrostatic repulsion between the microtoroid and nanorod surfaces, resulting in brief trapping of the nanorod followed by diffusion away from the resonator~\cite{Carousel}.  

In the right inset of Fig.~\ref{fig:data}b we show a histogram of the frequency shifts observed in the second experiment.  This histogram was found by first removing the constant frequency drift of -6.45 MHz/s and then identifying the size and location of step-like transitions using a step-finder algorithm~\cite{Little}.  As expected, frequency shifts with larger amplitudes were less common due to the strong dependence of the frequency shift on the orientation and position of the binding nanorod.  Out of the 35 events observed, the largest shift was -41.6 MHz (SNR of 12.2), which occured around 17 s.        

It is apparent from Fig.~\ref{fig:data} that the achieved SNR was dramatically different in the two experiments (by a factor of $\sim$ 4).  While the noise in the two experiments was comparable (3.2 MHz vs. 3.4 MHz), the difference in signal amplitudes accounts for the large improvement in SNR: i.e., the largest frequency shifts observed in the two experiments were -10.3 MHz and -41.6 MHz, respectively.  This factor of $\sim$ 4 can be explained by taking into account differences in microtoroid size and optical mode volume.  The smaller frequency shifts observed in the first experiment were likely due to poor mode volume, whereas in the second experiment we believe that the optical mode was close to the fundamental mode, as we will discuss later.

There are important differences between plasmonic nanoparticles and the dielectric nanoparticles detected in most WGM sensing experiments.  To elucidate this and examine whether the observed frequency shifts are consistent with nanorod binding events, we now calculate the expected frequency shifts for nanorods of the geometry used.  In the dipole approximation, the frequency shift due to a nanoparticle perturbing the electric field of a WGM resonator is given by~\cite{Arnold}
  
\begin{equation}
\frac{\delta \omega}{\omega} \simeq -\frac{ \textrm{Re}[\alpha] |E_0(\vec{r})|^2}{2 V |E_{0,\rm{max}}|^2}  
\label{eq:freq_shift}
\end{equation}
where $\alpha$ is the polarizability of the nanorod, $E_0(\vec{r})$ is the WGM electric field and $V$ is the optical mode volume of the resonator.  Nanorods can be well approximated as rotational ellipsoids with polarizability~\cite{Kreibig}
\begin{equation}
\alpha = \epsilon_b \frac{\epsilon(\omega)-\epsilon_b}{\epsilon_b+(\epsilon(\omega)-\epsilon_b)L}V_p
\label{eq:alpha}
\end{equation}
where the depolarization factor $L$ depends on the aspect ratio $R$ of the ellipsoid, $\epsilon(\omega)$ is the frequency-dependent dielectric function of Au~\cite{Johnson}, $\epsilon_b$ is the permittivity of water and $V_p$ is the volume of the nanoparticle.  

In Fig.~\ref{fig:nanorods}a, we compare the measured absorption spectrum from our nanorods with a theoretical calculation based on the dipole approximation above which averages over all orientations~\cite{Kreibig}.  In the measured spectrum (dotted curve), the nanorods exhibit a transverse resonance near 525 nm and a larger longitudinal resonance around 760 nm.  Spectra calculated in the dipole approximation are shown with a dashed curve for a single 39 nm$\times$10 nm nanorod and with a solid curve for a distribution of nanorods (average aspect ratio $\mu_R$ = 3.9 and standard deviation $\sigma_R$ = 0.39 taken from the manufacturer Sigma-Aldrich).  As can be seen, the dipole approximation gives excellent agreement with our measured spectrum over the wavelength range of interest.  However, in order to obtain a reliable estimate of the frequency shift, we must also know the optical mode volume $V$.  

Using SEM, we measured the major and minor diameters of the microtoroid used in the second experiment to be $D$ = 70 $\mu$m and $d$ = 6 $\mu$m, respectively.  We then used finite element modeling (COMSOL Multiphysics 3.4) to calculate a mode volume of $V$ = 350 $\mu$m$^3$ for the fundamental mode, for which the ratio $|E_0(\vec{r})|^2 / |E_{0,\rm{max}}|^2$ is about 0.3 at its equatorial maximum.  In Fig.~\ref{fig:nanorods}b we show the expected maximum frequency shift as a function of $R$ using Eq.~\ref{eq:freq_shift}-\ref{eq:alpha}.  For a nanorod with average aspect ratio ($R$ = 3.9), we expect a frequency shift of -47.5 MHz, slightly larger than the result found in our second experiment. This provides some evidence that the optical mode used in our second experiment was close to the fundamental mode, although this was not directly verified.  Additionally, the average aspect ratio of $R$ = 3.9 is consistent with the dimensions of the nanorods which we found on the microtoroid surface using SEM (see inset of Fig.~\ref{fig:nanorods}b for a representative image).  Interestingly, the sign of the expected frequency shift $\delta \omega$ depends on the detuning from the nanorod's plasmon resonance, with $\delta \omega > $ 0 for blue detuning.  This is not the case with dielectric nanoparticles.  On plasmon resonance, Re[$\alpha$] = 0 and therefore $\delta \omega = $ 0 through Eq.~\ref{eq:freq_shift}, which in our case happens for a nanorod with $R$ = 4.1 at $\lambda$ = 780 nm (Fig.~\ref{fig:nanorods}b).  In Fig.~\ref{fig:nanorods}c we show the size distribution of nanorods used in our experiments, where the shaded and non-shaded regions correspond to areas of expected negative and positive frequency shifts, respectively.  In our experiments, we did not find conclusive evidence of these positive frequency shifts, which may be due to the fact that the microtoroid surface red-shifts the resonant wavelength of the nanorod and therefore shifts the zero-crosing in Fig.~\ref{fig:nanorods}b to a larger value of $R$ where the probability of detection is much smaller (Fig.~\ref{fig:nanorods}c).  In addition, nanorods with Re[$\alpha$] $<$ 0 experience repulsive optical forces from the optical field~\cite{Pelton} and therefore should be difficult to detect in general.  

As shown in Ref.~\cite{Swaim}, plasmonic resonances in Au nanorods are expected give rise to large frequency shift enhancements and could enable single molecule detection under practical experimental conditions~\cite{Shopova, Santiago-Cordoba, Dantham}.  In Fig.~\ref{fig:nanorods}d, we show the calculated enhancement in $\delta \omega_m$ (single protein frequency shift) for a Au nanorod with $R = $ 3.9.  These calculations consider a protein with refractive index $n = $ 1.5, radius $a$ and polarizabilty $\alpha_m= 4 \pi \epsilon_b a^3 (\epsilon_m-\epsilon_b) / (\epsilon_m+2\epsilon_b)$, where $\epsilon_m=n^2$, that is bound to the tip of the nanorod.  The frequency shift induced by such a protein in a bare microtoroid $\delta \omega_0$ can be found using Eq.~\ref{eq:freq_shift}, and is shown with squares in Fig.~\ref{fig:nanorods}d.  The enhanced frequency shift $\delta \omega_R$, shown with circles in Fig.~\ref{fig:nanorods}d, is calculated as described in Ref.~\cite{Swaim}.  The inset of Fig.~\ref{fig:nanorods}d shows the enhancement factor as a function of protein radius $a$.  One can see that the plasmonic enhacnement is predicted to increase the expected frequency shift above our current noise floor ($\sim 2$ MHz for $\tau = $ 1 ms) for proteins with radii $a > $ 2 nm.  Under these conditions, this would lead to a minimum detectable protein radius of $a = $ 2 nm, which is smaller than the BSA protein.  

In summary, we have demonstrated optical detection of 39 nm$\times$10 nm gold nanorods using a microtoroid stabilized via the Pound-Drever-Hall method.  By volume, these nanoparticles are $\sim$3 smaller than the smallest dielectric spherical particles detected to date using the WGM sensing principle~\cite{Lu}, which shows that plasmonic resonances and large aspect ratios can significantly reduce the volume of detectable nanoparticles.  Similarly, we show that these properties lead to enhanced biosensing SNR, with a minimum detectable protein radius of $a$ = 2 nm for a nanorod with $R$ = 3.9.

We would like thank David Thompson for help with the absorption spectroscopy measurement, and the Australian National Fabrication Facility for use of their fabrication facilities.  

This research was funded by the Australian Research Council Grant No. DP0987146.


\end{document}